\documentclass[final]{elsarticle} 

\usepackage[english]{babel}
\usepackage[latin1]{inputenc}

\usepackage{amsmath}
\usepackage{amssymb}

\usepackage{subfigure}

\usepackage[nolist,nohyperlinks]{acronym}

\usepackage{lineno}

\biboptions{numbers,square,sort&compress}

\journal{Nucl. Instr. and Meth. A}

\begin{document}

\begin{frontmatter}



\title{Simulation of the BESIII Endcap Time of Flight Upgrade}


%0W7Q M-Ullrich
%6EFR Ulle15
%W9EF spam
%ICXU uni

\author[giessen]{M. Ullrich\fnref{correspondingauthor} }
\author[giessen]{W. K\"uhn}
\author[giessen]{Y. Liang}
\author[giessen]{B. Spruck}
\author[giessen]{M. Werner}

\address[giessen]{2. Physikalisches Institut, Justus-Liebig-Universit\"at Gie{\ss}en, Germany}
 \fntext[correspondingauthor]{Corresponding Author}

\begin{abstract}
The results of a full simulation of an endcap \acl{ToF} detector upgrade based on \aclp{MRPC} for the \acs{BESIII} experiment are
presented. The simulation and reconstruction software is based on Geant4 and has been implemented into the \acl{BOSS}.
The results of the simulations are compared with beam test results and it is shown that a total time resolution $\sigma$ of about 80~ps can be achieved 
allowing for a pion and kaon separation up to momenta of 1.4~GeV/c at a 95\% confidence level.
\end{abstract}

\end{frontmatter}

%\linenumbers

\section{Introduction}
\label{S:introduction}
To enhance the capability of particle identification, it is planned to replace the actual endcap \ac{ToF} system of the \ac{BESIII} 
experiment by a system based on \acp{MRPC} in summer 2015. \acp{MRPC} are gas based avalanche detectors and are already used in various
high energy physics experiments \cite{mrpc_hades,mrpc_alice,mrpc_star}. They are characterized by a high time resolution and a large detection efficiency and can be produced
at low cost with a high granularity \cite{mrpc_production_1,mrpc_production_2}. Details about the \ac{MRPC}'s operational principle are given in \cite{mrpc_history_1,mrpc_history_2}.\\
This paper presents the simulation and reconstruction software developed for the \ac{MRPC} upgrade at the \ac{BESIII} experiment. The software is implemented into the \ac{BOSS} \cite{boss_system}, capable 
to perform a full simulation and reconstruction of the \ac{BESIII} detector. Besides a comparison of the simulated results with the data taken during a beam test \cite{beamtest_ihep},
 the enhanced capability of pion and kaon separation in the upgraded \ac{BESIII} detector is investigated.

\section{BESIII Experiment}
\label{S:Experiment}
The \ac{BESIII} experiment \cite{bes3_detector} is operated at the \ac{BEPCII} at the \ac{IHEP}. The symmetric $e^+e^-$ experiment
covers the center-of-mass energy range $\sqrt{s}$ from 1.8~GeV to 4.6~GeV and is optimized for the investigation of $\tau$ and charm 
physics.
The double ring $e^+e^-$ collider is designed for a peak luminosity of $10^{33}\text{ cm}^{-2}\text{ s}^{-1}$ at beam currents of 0.93~A
at $\sqrt{s}=3.773$~GeV.

The detector consists of 5 major components:\\
(1) The innermost component is a helium gas based \ac{MDC} with 43 layers in total, a single wire spatial resolution of 135 micron and an
angular acceptance of about 93\% of 4$\pi$. The $dE/dx$ resolution is better than 6\% and the momentum resolution for charged particles
with momenta of 1.5~GeV/c in a 1~Tesla magnetic field is 0.5\%.\\
(2) The next outer detector is a \ac{ToF} system used for particle identification. It is composed of a two-layer-structure of 2300~mm long BC-408 plastic scintillators in the barrel region 
($|\cos\theta < 0.82|$, with $\theta$ being the polar angle) and of a one-layer-structure made of trapezoidal shaped BC-404 scintillators with a height of 480~mm in the endcap regions ($0.84 < |\cos\theta| < 0.95$).
The time resolution $\sigma$ for Bhabha scattering events was measured to be 78~ps in the barrel region, but only about 150~ps in the endcap region \cite{bes3_tof_endcap_calibration}. The worse resolution in the endcaps
is mainly caused by multiple scattering interactions in the endcap plate of the \ac{MDC} and material located between \ac{MDC} and endcap \ac{ToF} systems, such as electronics and cables \cite{bes3_tof_endcap_scattering}. 
The resolution for pions and kaons in the endcaps is about 135~ps and thus 
allowing for a pion/kaon separation up to momenta of about 1~GeV/c \cite{bes3_tof_endcap_calibration}.\\
(3) The \ac{EMC} surrounds the \ac{MDC} and \ac{ToF} systems and consists of 6240 thallium doped cesium-iodide crystals. It provides an energy resolution of 2.5\% (5.0\%) and a 
position resolution of 6~mm (9~mm) for photons with an energy of 1~GeV in the barrel (endcap) parts.\\
(4) The three inner detector systems are surrounded by a superconducting solenoid magnet, which provides an axial uniform magnetic field of 1.0~Tesla.\\
(5) The outermost detector is a muon chamber system, which is embedded in the flux return of the magnet. It consists of 8 (9) layers of \ac{RPC} in the barrel (endcap) region
and provides a spatial resolution of 2~cm.

\section{Endcap \ac{ToF} Upgrade}
\label{S:MRPC_Upgrade}
\begin{figure}
\centering
\includegraphics[width=0.95\linewidth]{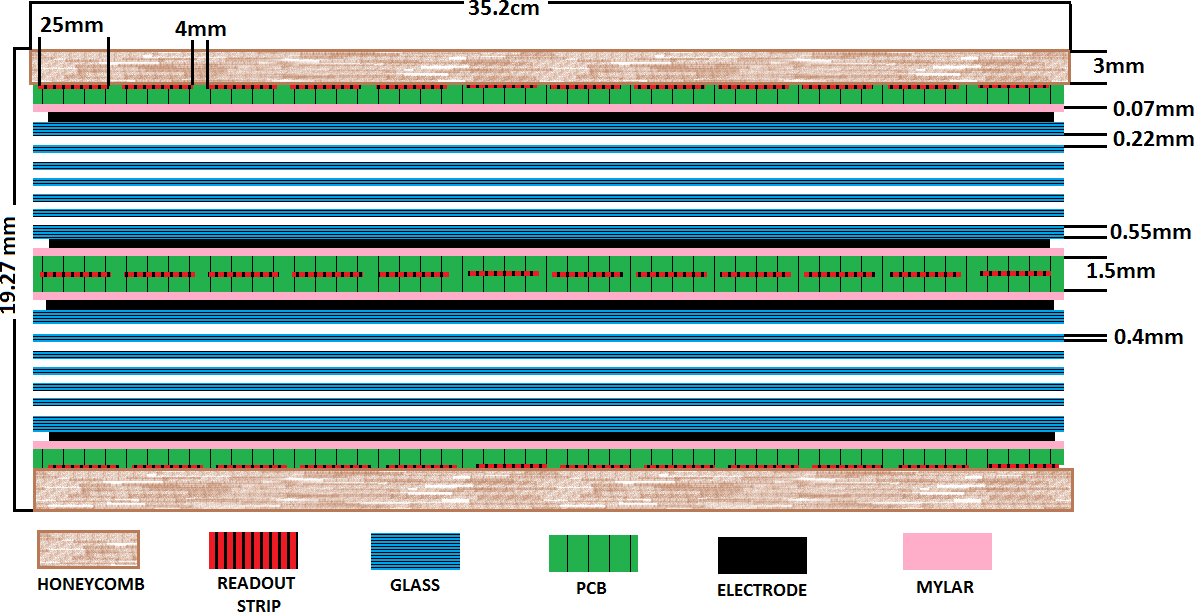}\\ \ \\
\includegraphics[width=0.45\linewidth]{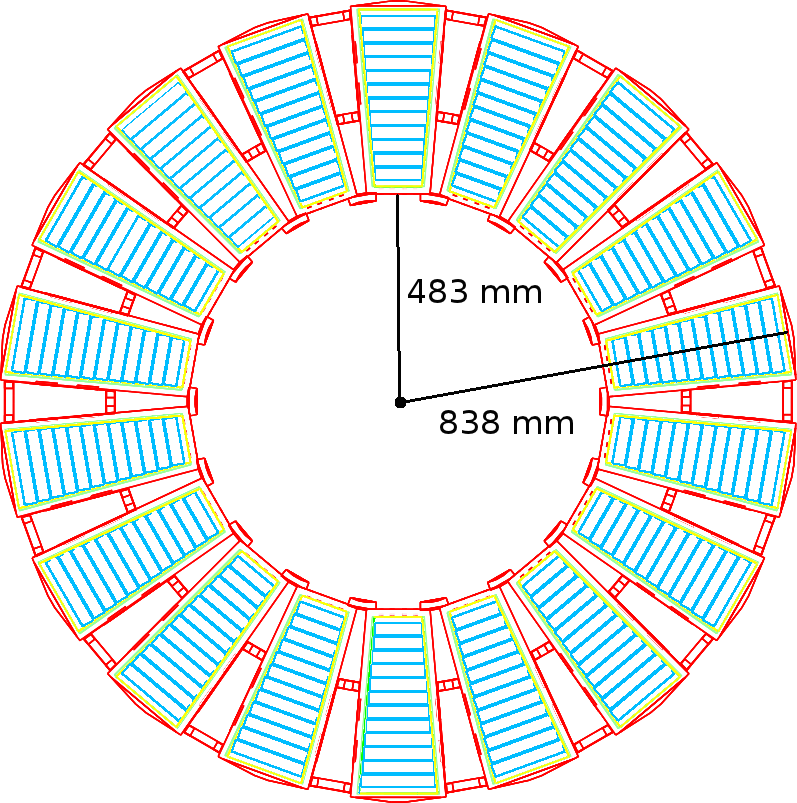}
\includegraphics[width=0.27\linewidth]{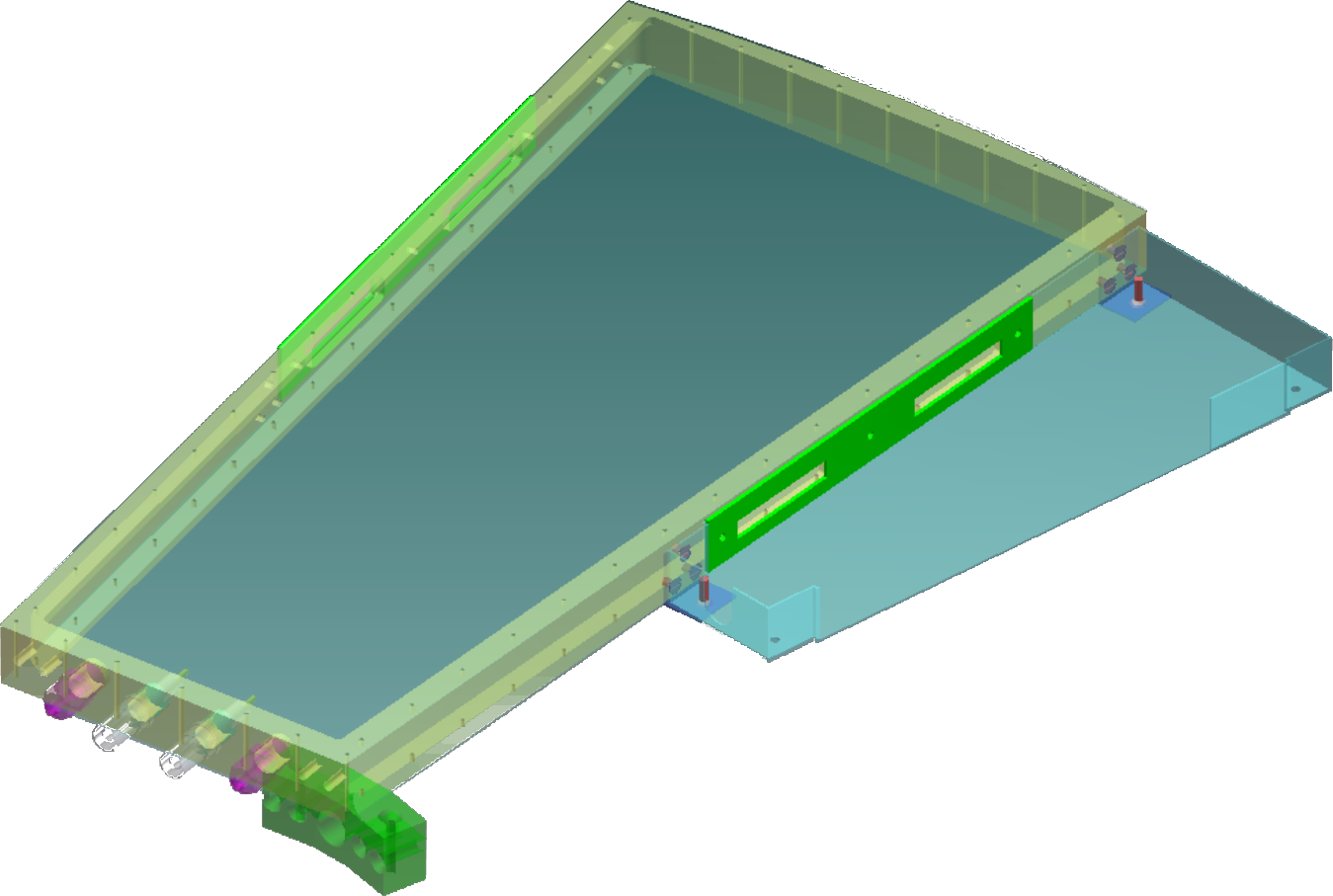}
\includegraphics[width=0.26\textwidth]{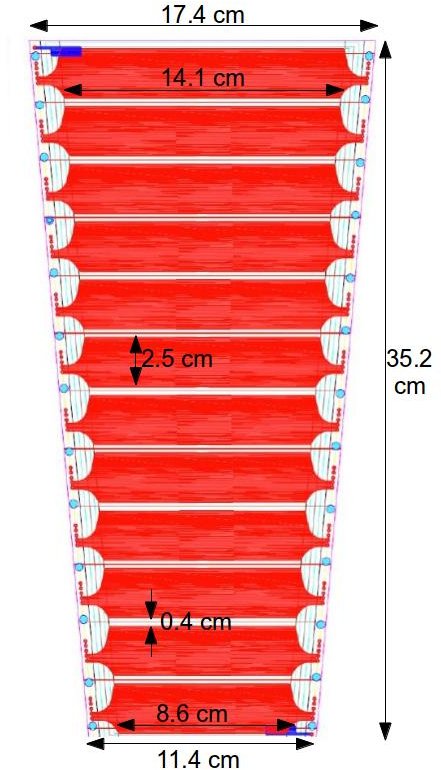}
\caption{Top: Sketch of the \ac{MRPC} module planned to be used for the endcap \ac{ToF} upgrade at the \ac{BESIII} experiment. Bottom left: Arrangement of a single layer of \ac{MRPC} modules. 
Bottom middle: The aluminum frame and the box for the \ac{FEE}. Bottom right: Top view and dimensions of the readout strips of a MRPC module.}
\label{fig:sideview_mrpc} 
\end{figure}

The \ac{MRPC} modules planned to be used for the endcap \ac{ToF} upgrade at the \ac{BESIII} detector are double stack \acp{MRPC} 
with in total 12 gas layers with a thickness of 0.22~mm, separated by 0.4~mm thick glass plates. The glass plates covering the electrodes have a thickness of 0.55~mm. 
The electrodes are separated by 0.07 mm thick mylar layers from the printed circuit board containing the readout strips with a characteristic impedance of less than 50~$\Omega$.
The module is covered with a honeycomb structure on both sides for protection against damages. Figure \ref{fig:sideview_mrpc} shows a sketch of a \ac{MRPC} module and a top view 
and the dimensions of the readout strips.\\
The modules will be operated at a voltage of about 14~kV (corresponding to an electric field strength of about 106~kV/cm) 
with a gas mixture of 90\%~C$_2$F$_4$H$_2$ (Freon), 5\%~SF$_6$ and 5\% iso-butane. Each module will contain 12 readout strips, which are read out on both sides. Thus, the differences
in the arrival time of the signals can be used to reconstruct a one-dimensional impact position of a particle.
The modules will be placed in a 25~mm thick aluminum frame on one side of which an additional box is mounted containing the \ac{FEE} (figure \ref{fig:sideview_mrpc}).
 To overcome the dead area arising due to the additional box for the \ac{FEE} and the dead area at the borders of a single module, each endcap will consist of two layers of modules, which are shifted towards each other.
A single layer will contain 18 modules arranged in a ring (figure \ref{fig:sideview_mrpc})  and will thus have 432 readout channels, which are read out by NINO chips \cite{nino_1}.
The rising edge of the NINO chip's signal encodes the time measurement and the length of the pulse (time over threshold) encodes the charge induced in the readout strips. The signals produced by
the NINO chips will be fed into \acp{HPTDC} \cite{hptdc_paper} providing a time resolution of 25~ps.

\section{Simulation}

The geometric and material properties of the \ac{MRPC} detector upgrade are implemented into \ac{BOSS} \cite{boss_system} by means of Geant4's concept of logical and physical volumes \cite{geant4}. The Geant4 physics list
used for the simulations is the standard list QGSP\_BERT\_CHIPS.

The signal electron multiplication in the simulation is described using statistical language. Detailed information about electron multiplication 
in gases at high electric fields can be found in \cite{avalanche_legler}, whereas here only a brief introduction 
is given. Starting with a single electron at $x=0$ the probability of having $n$ electrons at $x+dx$ can be calculated (in first order) according to
  \begin{equation}
	P(n)=\begin{cases}
	k\frac{\bar{n}(dx)-1}{\bar{n}(dx)-k} ~ ~ ~ ~ \qquad  \qquad  \qquad  \qquad  \qquad & \text{for } n=0\\
	\bar{n}(dx) \left(\frac{1-k}{\bar{n}(dx)-k}\right)^2 \left(\frac{\bar{n}(dx)-1}{\bar{n}(dx)-k}\right)^{n-1} ~ ~ ~ & \text{for } n>0
       \end{cases}
      \label{eq:probability_electrons}
 \end{equation}
with  $\bar{n}(dx)=exp((\alpha-\eta)dx)$, $k=\eta/\alpha$, $\alpha$ the Townsend coefficient and $\eta$ the attachment coefficient \cite{avalanche_legler}. Equation \ref{eq:probability_electrons}
can only be employed for $\alpha\neq\eta$ and $\alpha\neq0$, which holds for the operation of a \ac{MRPC} in a high electric field.\\
The simulation of the electron avalanche is based on the 1D-model introduced by \textit{Riegler, Lippmann and Veenhof} \cite{avalanche_simulation_1d}: At the beginning each gas layer of the detector is divided into $N$ steps of size $dx$ 
and the primary electron-ion pairs are distributed along the gap according to the information provided by Geant4. For each single electron, the number of electrons at 
a distance $x+dx$ is calculated according to the probability 
distribution of equation \ref{eq:probability_electrons}. The latter step is repeated until all electrons have left the gap.\\
To speed up the simulation process, for sufficiently large numbers of electrons ($\approx150$) the central limit theorem is applied. The number of electrons in the next step is simply calculated by drawing a 
random number from a Gaussian distribution with mean ${\mu=n(x)\bar{n}(x)}$ and standard deviation $\sigma=\sqrt{n(x)}\hat{\sigma}(dx))$ with $\hat{\sigma}(x)^2=\frac{1+k}{1-k}\bar{n}(x)(\bar{n}(x)-1)$ \cite{weighting_field_mrpc}.\\
However, this simple model has a serious drawback as it does not consider space charge effects and thus results in a steady growth of the avalanche charge, which is not observed in experiments. To take into account
space charge effects a simple cut-off is applied \cite{Lippmann_space_charge_effects_rpc}. The avalanche is saturated and does not grow anymore as soon as the number of charge carriers exceeds $1.5\cdot10^{7}$. 
This simple cut-off does not affect the main characteristics of the \ac{MRPC}, such as efficiency and resolution, as they are only sensitive to the first stage of the avalanche \cite{space_charge_avalanche}.\\
The current induced in the readout strips by the moving charge carriers is calculated utilizing \textit{Ramo's} theorem $I(t)=E_w v_d Q N(t)$ \cite{ramos_theorem}, with $v_d$ being the drift velocity and $Q$ the charge of
the charge carriers. $E_w$ is the electric weighting field~\cite{weighting_field_mrpc} and $N(t)$ the number of charge carriers at a certain time~$t$. 

\begin{figure}
\centering
\includegraphics[width=0.80\textwidth]{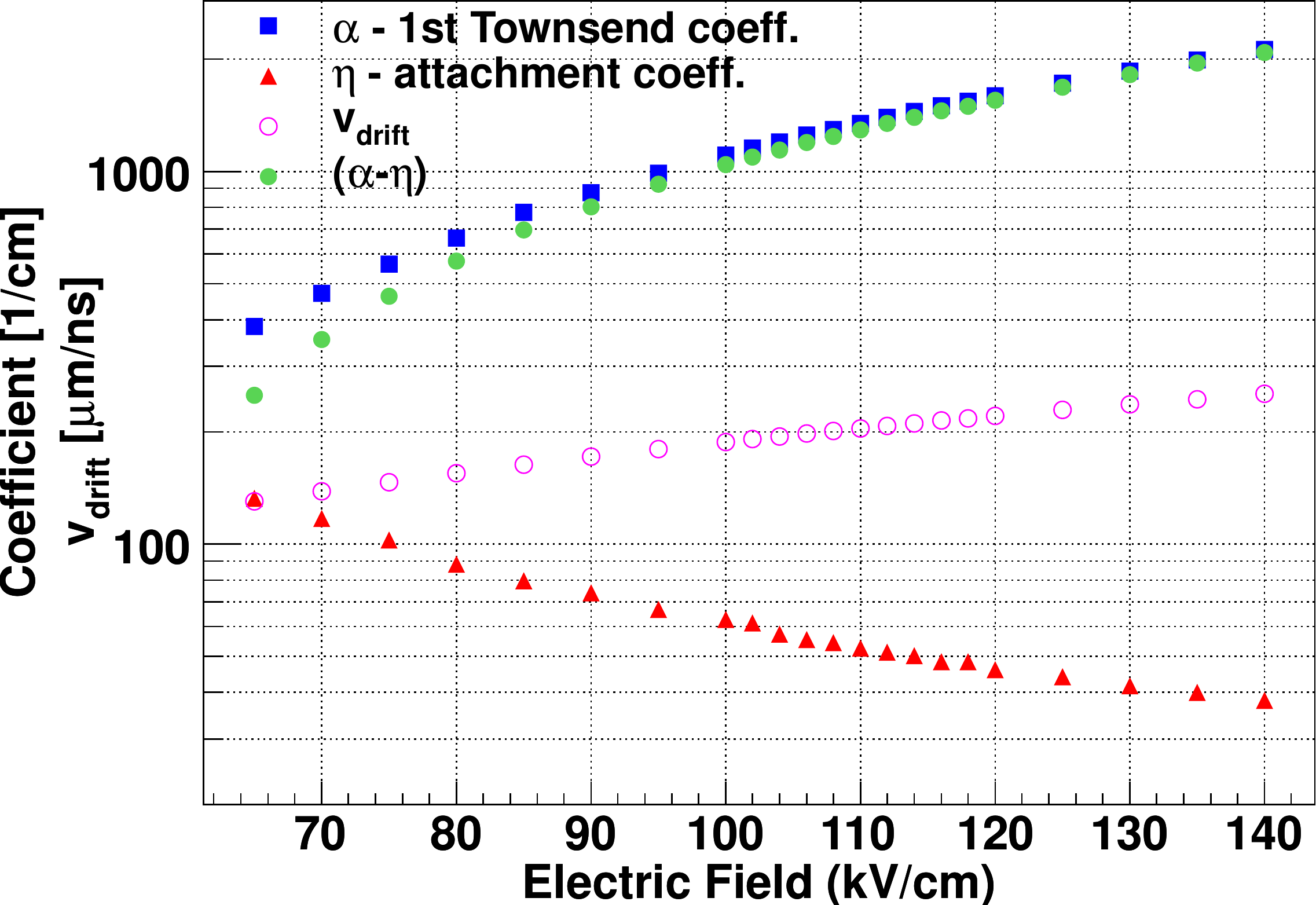}\label{fig:magboltz_simulation_1}
\caption{Simulation results from MAGBOLTZ for a gas mixture of 90\%~C$_2$F$_4$H$_2$ (Freon), 5\%~SF$_6$ and 5\% iso-butane at standard conditions for temperature and pressure. 
	    First Townsend coefficient $\alpha$ (blue squares), attachment coefficient $\eta$ (red triangles), effective Townsend coefficient $\alpha - \eta$ (green dots) and 
                        the avalanche's drift velocity v$_{\text{drift}}$ (hollow purple circles).
	   }
\label{fig:magboltz_simulation}
\end{figure}

The Townsend coefficient $\alpha$, the attachment coefficient $\eta$ and the drift velocity $v_d$ of the charge carriers are required to perform the electron multiplication and to determine
the induced current in the readout strips. These parameters are not provided by Geant4 and thus are calculated with MAGBOLTZ \cite{magboltz_1}. The results for the gas mixture of  90\%~C$_2$F$_4$H$_2$ (Freon),
5\%~SF$_6$ and 5\% iso-butane at standard conditions for pressure and temperature for different electric fields strengths are shown in figure~\ref{fig:magboltz_simulation_1}.\\
%Figure~\ref{fig:magboltz_simulation_2} also shows the longitudinal and the transversal diffusion coefficients, which are ignored in this paper as they are not required by this simple model.\\ 
The average energy deposition required to produce an ionization (an electron-ion pair) in the gas mixture is also required at simulation level by Geant4. It was set to 40~eV for all simulations.
\begin{figure}
  \subfigure[]{\includegraphics[width=0.32\textwidth]{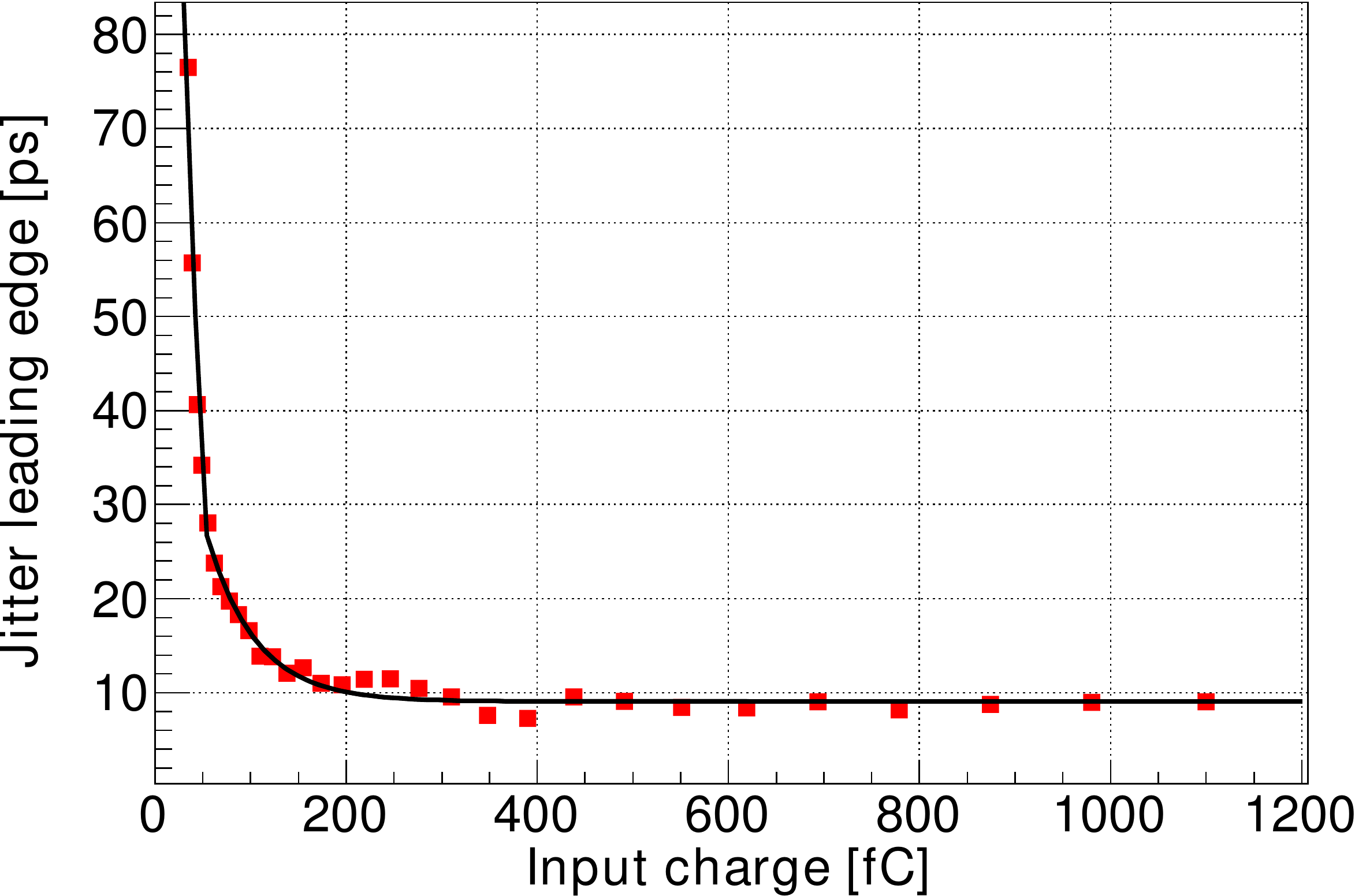}\label{fig:jitter_leadingedge}}
  \subfigure[]{\includegraphics[width=0.32\textwidth]{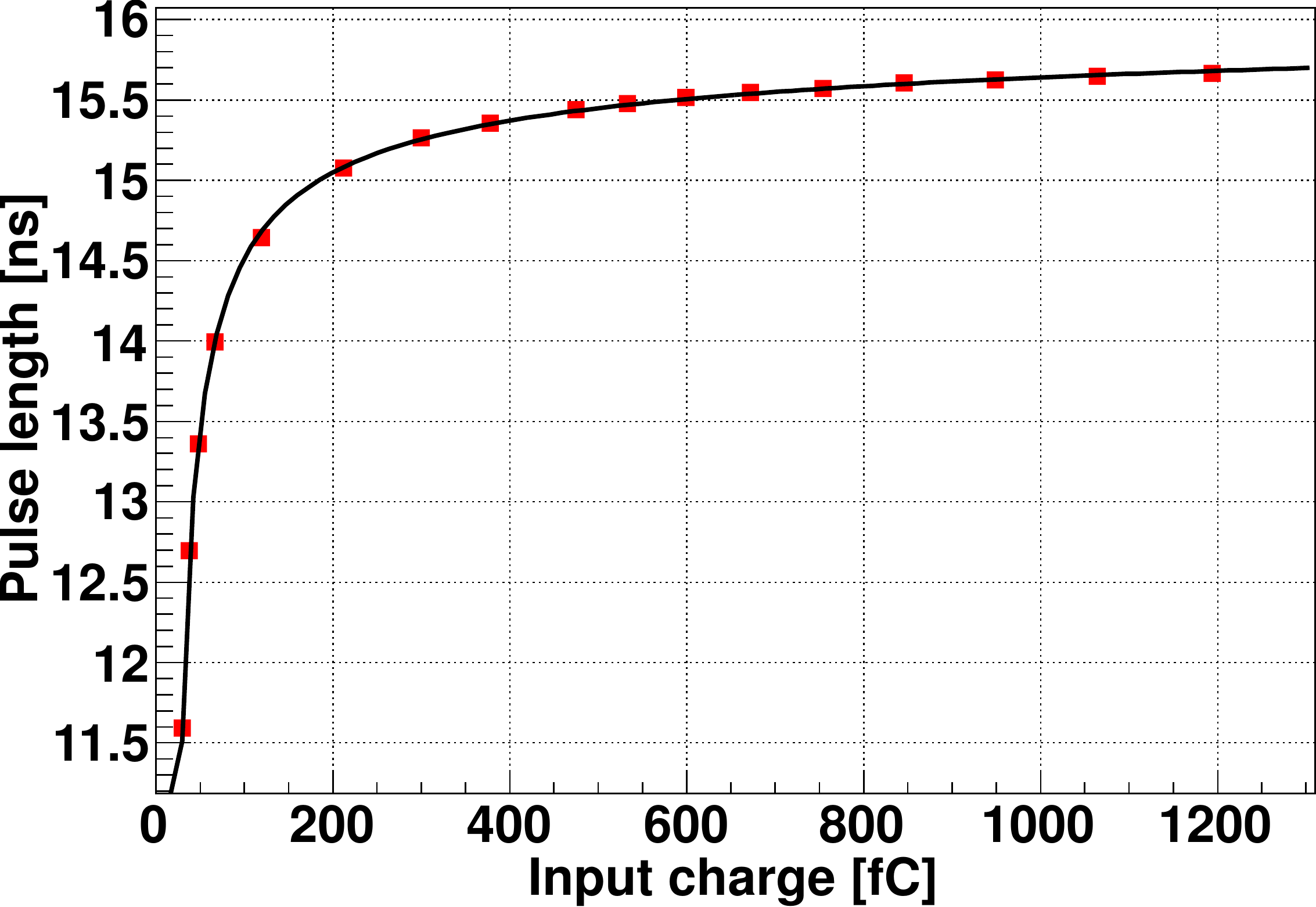}\label{fig:pulselength_vs_inputcharge}}
  \subfigure[]{\includegraphics[width=0.32\textwidth]{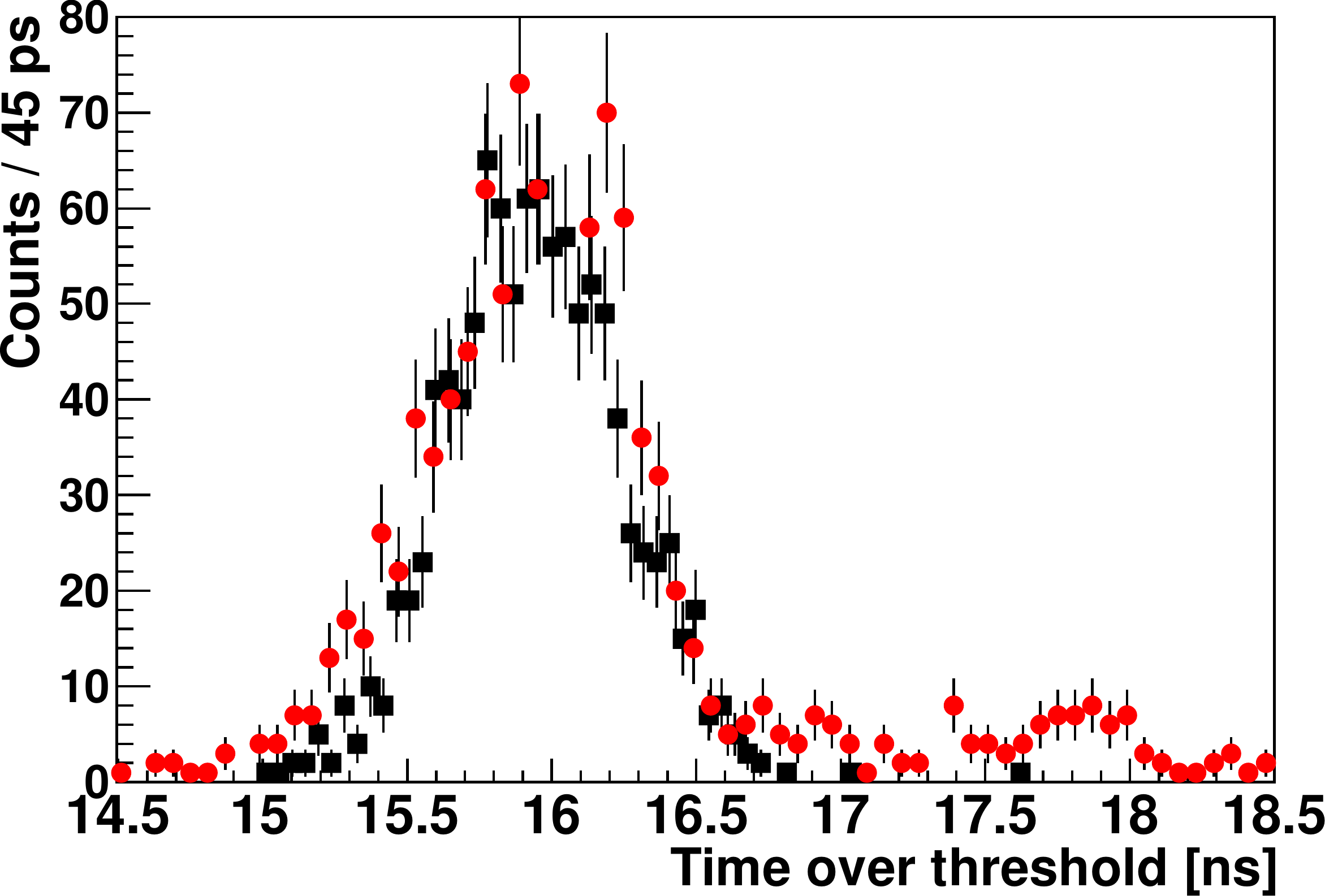}\label{fig:charge_spectra}}
  \caption{\textbf{(a)} Input charge into the NINO chip versus the jitter of the leading edge \cite{private_communication_daihongliang}. The measurement was performed using the HP8082 signal generator as signal source and
the Tektronix DSA70404 oscilloscope as a probe. \textbf{(b)} Input charge into the NINO chip versus the produced pulse length \cite{private_communication_daihongliang}. \textbf{(c)} Comparison of simulation results (black squares) and measurements (red dots) for the time over threshold.}
\label{fig:simulation_parameter_1}
\end{figure}

Uncertainties in the time domain arising from the avalanche process due to fluctuations in the electron multiplication are covered by the model and are about 25~ps for the signal from all 12 gaps at an electric field 
strength of 110~kV/cm. 
Further uncertainties in the time measurement arise from the TDC resolution (25~ps), the jitter of the leading edge of the NINO chips, which depends on the input charge (figure \ref{fig:jitter_leadingedge}),
and additional components (such as cable or noise) chosen empirically to be $23$~ps. These components are considered by smearing the measured time value with a Gaussian distribution 
with  the corresponding standard deviation $\sigma$.
 
\begin{figure}
  \subfigure[]{\includegraphics[width=0.5\textwidth]{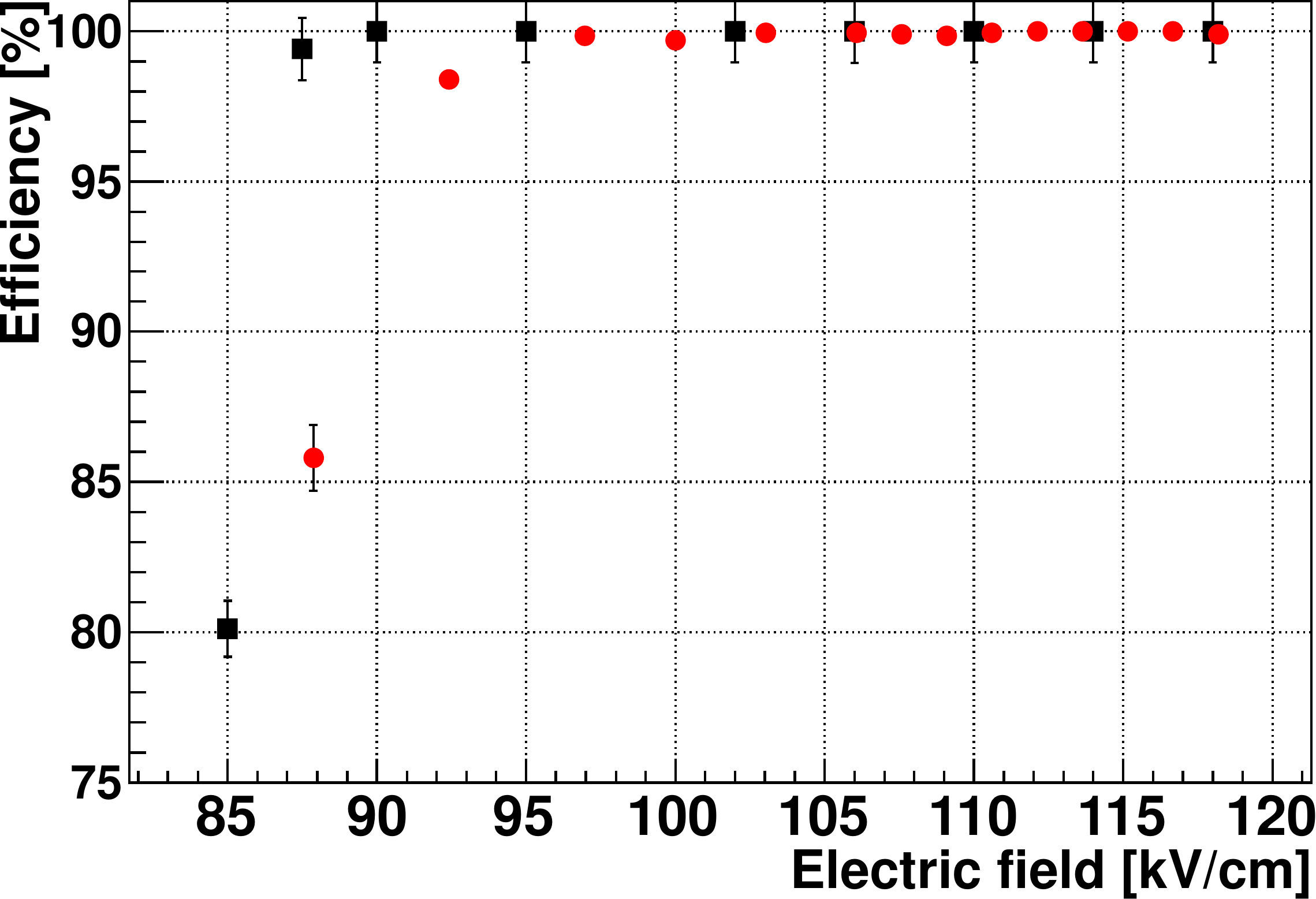}\label{fig:efficiency}}
  \subfigure[]{\includegraphics[width=0.5\textwidth]{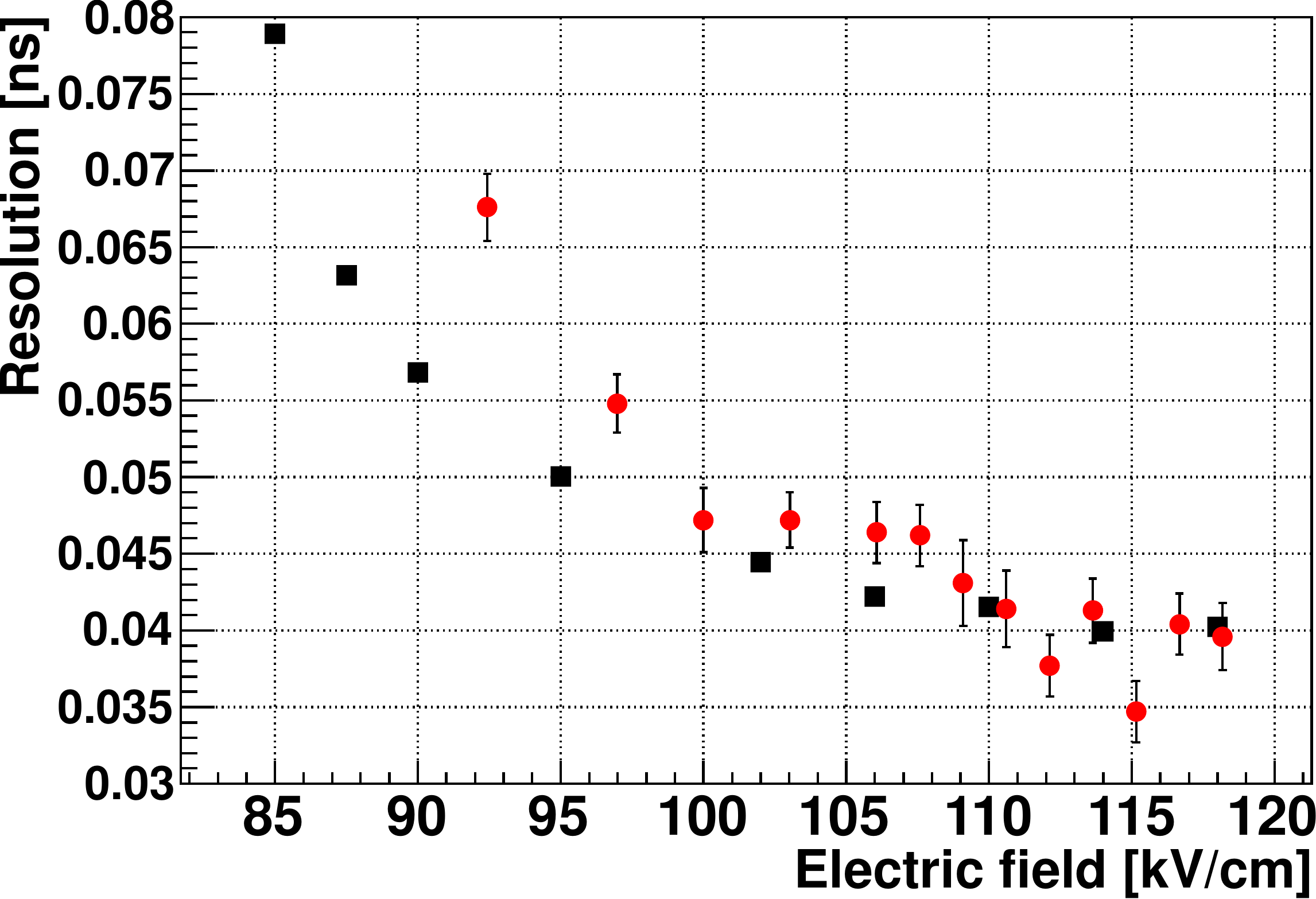}\label{fig:resolution}}
   \caption{\textbf{(a)} Comparison of simulated (black squares) and measured (red dots) efficiencies for different electric field strengths. 
	    \textbf{(b)} Comparison of simulated (black squares) and measured (red dots) time resolutions for different electric field strengths.}
\label{fig:simulation_result_1}
\end{figure}

Figure \ref{fig:simulation_result_1} shows a comparison of simulated and measured efficiencies and time resolutions for different electric field strengths. The measured results are from   
a \ac{MRPC} prototype test at the \ac{BEPCII} E3 beam line with protons with a momentum of about 600~MeV/c \cite{beamtest_ihep}. The threshold for signal detection was set to a charge induction of 
70~fC for a readout strip.
The simulated data agrees well with the measurements for a field strength larger than $\sim$100~kV/cm. For a field strength smaller than this value the time resolution is underestimated in the simulation.
%In this area recombination effects in the early stage of the avalanche may play a role, which are not considered in the model.
\\
The input charge simulated in the model is converted into the time over threshold measurement of the NINO chip by a function relating the input charge to the time over threshold measurement. 
The resulting converted pulse length has been then smeared with a Gaussian with a standard deviation of 0.3~ns, corresponding to an uncertainty of about 2\% within the conversion.
 The function to relate input charge and time over threshold has been generated 
with a pulser (figure \ref{fig:pulselength_vs_inputcharge}).
The result of the simulation is shown in figure \ref{fig:charge_spectra} and fits the measured time over threshold spectra well, except for values larger than $\sim$17~ns. 
This part of the spectra may arise from streamers, which are not covered by the avalanche simulation in the underlying model.

\section{Reconstruction}
The reconstruction software is also embedded into \ac{BOSS} and its tasks are to match the tracks reconstructed in the \ac{MDC} with the corresponding \ac{MRPC} signals, to apply walk corrections
and to correct for effects from electronics and cable delays, and to determine the transition time of the signals within the readout strips, which is required to improve the measured time information.

To match the reconstructed \ac{MDC} tracks with the \ac{MRPC} signals, the tracks are extrapolated to the outer sub-detector systems by means of a Geant4 based tracking and stepping algorithm, which considers 
magnetic deflection and ionization energy loss \cite{boss_system}. 
Around the extrapolated impact position on the \ac{MRPC} detector (which are all readout strips of the \ac{MRPC} module the extrapolated track is pointing to
and the readout strips of the directly adjacent \ac{MRPC} modules) a search is done for signals caused by the corresponding \ac{MDC} track. If two or more candidates exist, the strip with the largest charge deposition is chosen.\\ 
The transition time of the signal within the readout strip is reconstructed using the impact position of the extrapolated \ac{MDC} track. Figure \ref{fig:trans} shows the difference of the simulated and 
reconstructed transition time and has been fitted with a double Gaussian. The standard deviation of the first Gaussian (covering 87\% of the signals) is 14~ps (corresponding to a position resolution 
of about 0.4~cm), whereas the standard deviation of the second Gaussian is 34~ps. An explanation of the origin of the two different Gaussian components 
can not be given yet, but it may be related to the Geant4 based algorithm, which is used to extrapolate the impact position on the readout strip. If the extrapolated \ac{MDC} track does not point to a readout strip which has caused the signal
(for example the \ac{MDC} track is extrapolated to a strip, which does not have the largest charge deposition, but the neighbouring strip), the impact position (transition time) is calculated using the different
time of arrival of the signal on both sides of the readout strip. Here the achieved resolution is about 38~ps.

\begin{figure}
  \subfigure[]{\includegraphics[width=0.5\textwidth]{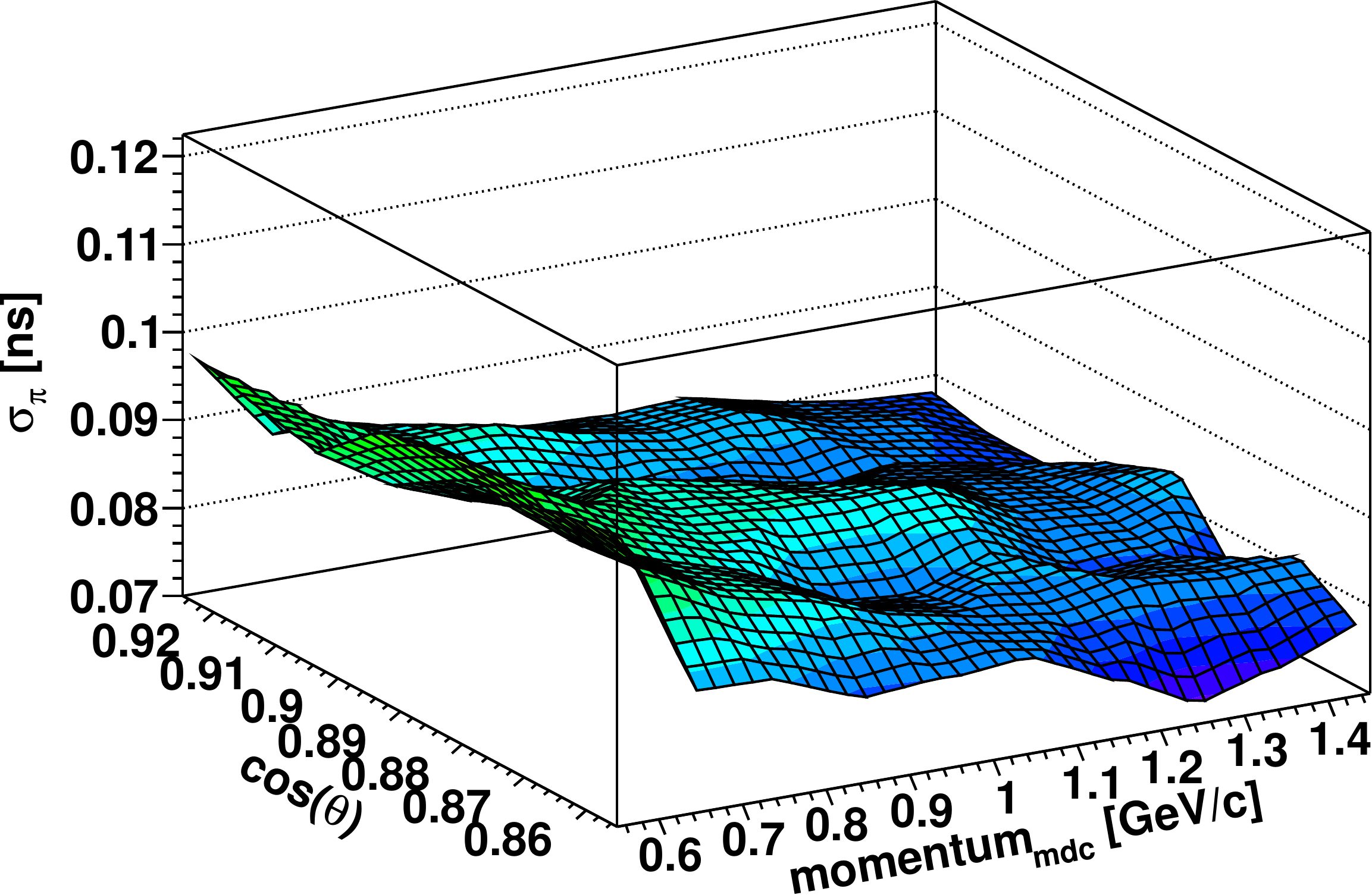}\label{fig:resolution_pion}}
  \subfigure[]{\includegraphics[width=0.5\textwidth]{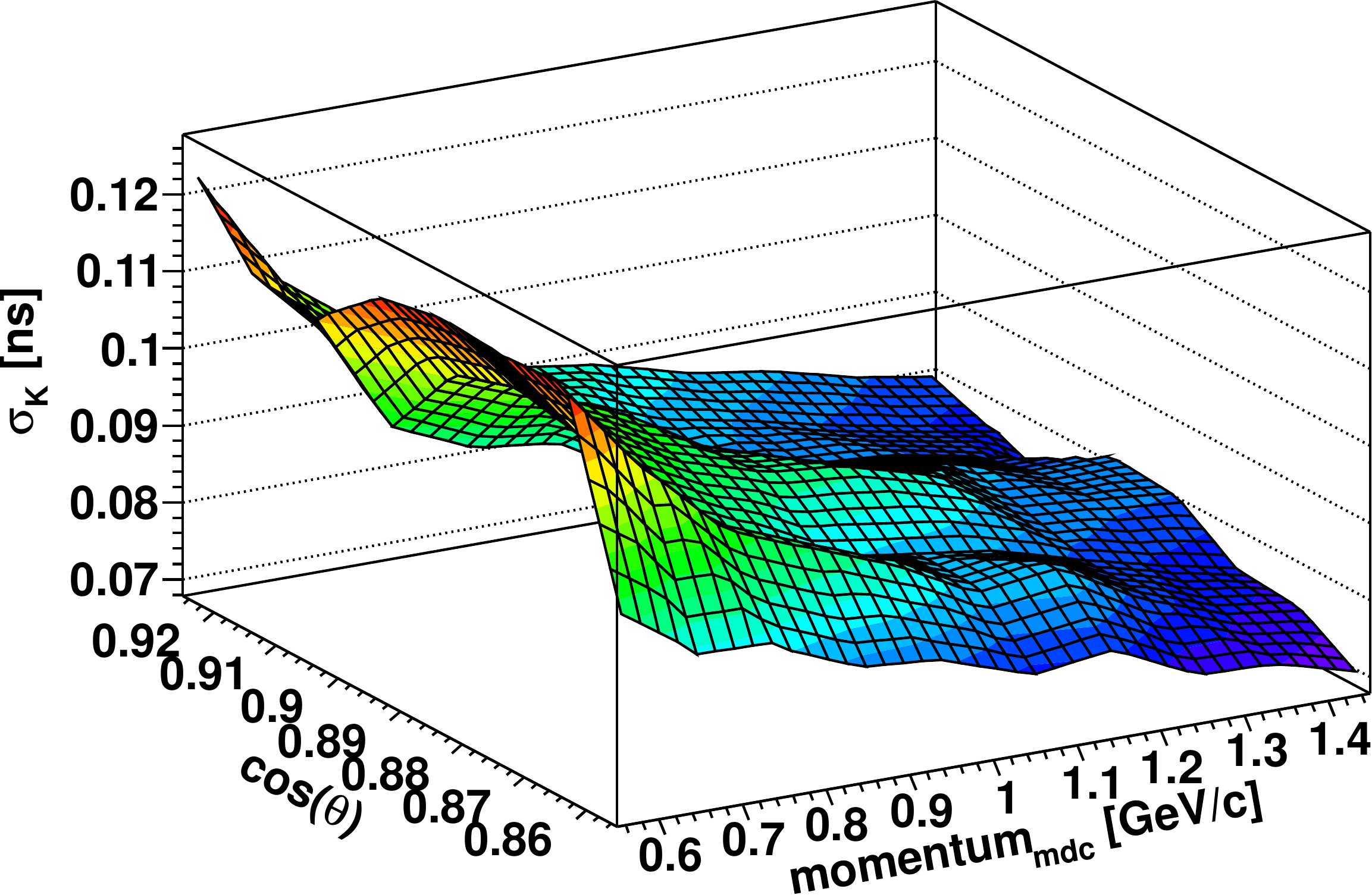}\label{fig:resolution_kaon}}
   \caption{Simulated resolution of the differences of measured and expected time of flight for pions \textbf{(a)} and kaons \textbf{(b)} in dependence of $\cos\theta$ and the momentum of the particle.}
\label{fig:resolution_kaon_and_pion}
\end{figure}

\section{Improvement in Pion and Kaon Separation}
\begin{figure}
  \subfigure[]{\includegraphics[width=0.5\textwidth]{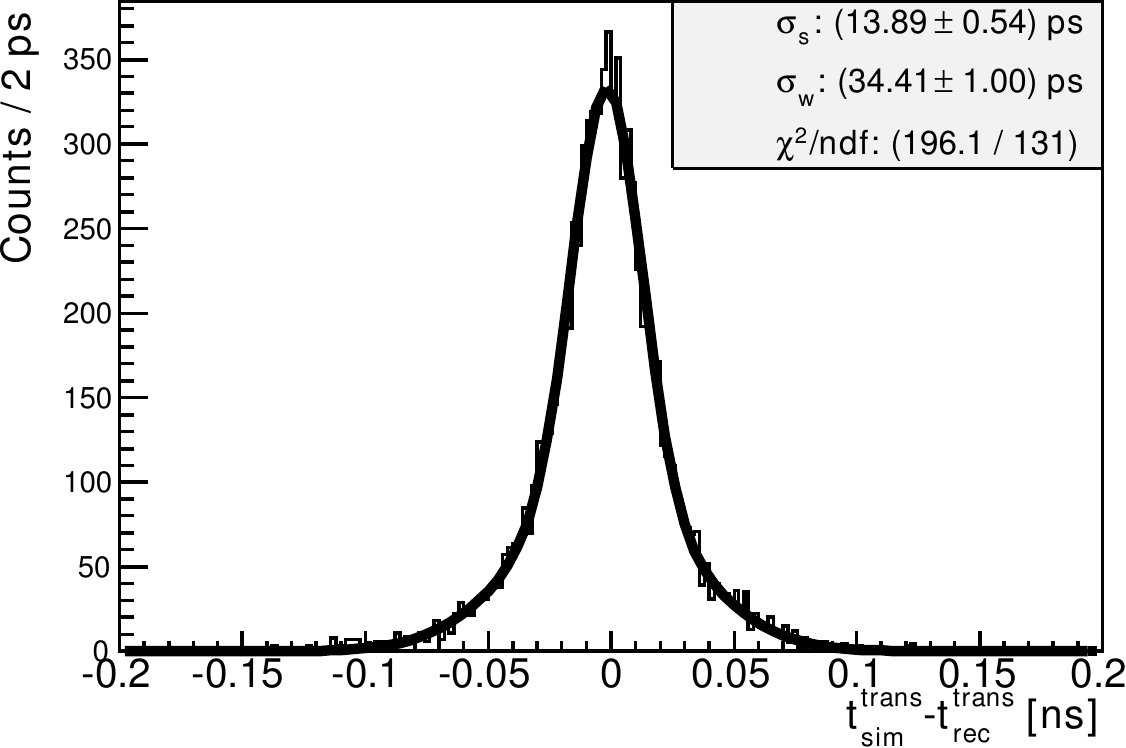}\label{fig:trans}}
  \subfigure[]{\includegraphics[width=0.5\textwidth]{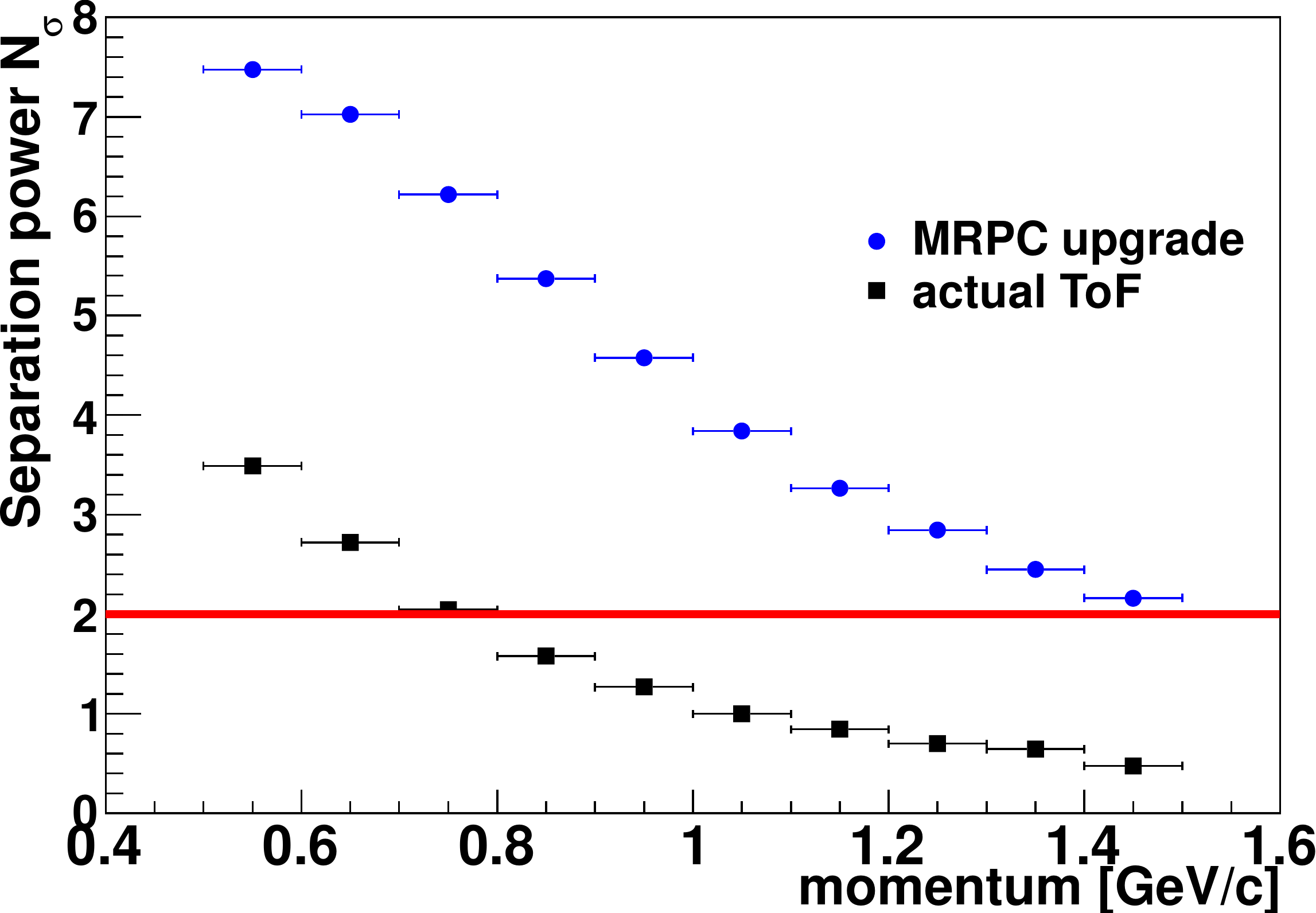}\label{fig:improvement}}
   \caption{\textbf{(a)} Difference of simulated and reconstructed signal transition time in a readout strip. \textbf{(b)} Comparison of the pion and kaon separation power for the simulated \ac{MRPC} upgrade (blue dots)
and the actual endcap time of flight detector (black squares).}
\label{fig:trans_and_improvement}
\end{figure}

One of the major tasks of the \ac{ToF} system at the \ac{BESIII} detector is to allow for the separation of pions and kaons with momenta larger than about 600~MeV/c
(starting from this momentum the difference in specific ionization energy loss of the particles can not be used anymore for their clear separation \cite{bes_physics_book}).\\
To study the improvement in pion and kaon separation, single particle events of pions and kaons with momenta between 0.5~GeV/c and 1.5~GeV/c have been simulated and fully 
reconstructed using the current \ac{ToF} system and the upgraded \ac{ToF} system in \ac{BOSS}. The pion and kaon separation power can be expressed as 
\begin{equation}
N_{\sigma}=\frac{|x_K-x_{\pi}|}{\sqrt{\sigma^2_K + \sigma_{\pi}^2}} \qquad ,
\end{equation}
with $x_i$ being the fitted difference of measured time of flight $t_{\text{meas.}}$ and expected time of flight $t_{\text{exp.}}$ and $\sigma_i$ the corresponding width of the Gaussian distribution. A separation power of $N_{\sigma}=2$ corresponds
to a misidentification probability of 7.9\%, a separation at 95\% confidence level to $N_{\sigma}=2.33$. The expected time of flight can be calculated according to 
\begin{equation}
t_{\text{exp.}}=\frac{L}{\beta c} = \frac{L}{ \frac{p/m}{\sqrt{1+p^2/m^2}} \cdot c},
\end{equation}
with $\beta$ being the relativistic velocity of the particle, $c$ the velocity of light, $p$ the momentum of the  particle measured in the \ac{MDC} and $m$ the mass of the particle. 
$L$ is the distance traveled by the particle (extrapolated track length), which has been determined by a separate reconstruction algorithm.\\
Using the simulated data the distributions $x_K$ and $x_{\pi}$ have been fitted with a Gaussian for different momentum intervals 
and the separation power $N_{\sigma}$ has been calculated. The width $\sigma$ of the distributions $x_K$ and $x_{\pi}$ is shown in figure \ref{fig:resolution_kaon_and_pion} for different 
momenta and $\cos\theta$ values. $\sigma$ rises for smaller momenta, as the resolution of the algorithm determining the track length worsens.\\
The separation power $N_{\sigma}$ for different momenta of the current \ac{ToF} system and the upgraded one is shown in figure \ref{fig:improvement}.
The improvement using the \ac{MRPC} based version is clearly visible. The upgraded system will allow for a pion and kaon separation up to momenta of 1.4~GeV/c at a 95\% confidence level.

\section{Summary and Conclusions}
The endcap \ac{ToF} system of the \ac{BESIII} experiment is planned to be upgraded using nowadays available \ac{MRPC} technology in summer 2015.
The implemented model for detector simulation is able to reproduce the beam test results and is fully implemented into the software framework 
of the \ac{BESIII} experiment. The reconstruction software is also embedded into the framework and allows for matching reconstructed 
\ac{MDC} tracks with the measurements of the \ac{MRPC} system. The impact position reconstruction allows for a correction of the time measurement
and so for reducing uncertainties arising due to the size of the readout strips.\\
The results of the full simulation and reconstruction show that a significantly improved pion and kaon identification at 95\% confidence level up to momenta of 1.4~GeV/c can be achieved.

\bibliographystyle{elsarticle-num}
\bibliography{paper_mrpc}

 \begin{acronym}[SQL]

  \acro{BESIII}{Beijing Electron Spectrometer III}
  \acro{BEPCII}{Beijing Electron Positron Collider II}
  \acro{IHEP}{Institute of High Energy Physics}

  \acro{BOSS}{BESIII Offline Software System}

  \acro{MDC}{Main Drift Chamber}
  \acro{ToF}{time of flight}
  \acro{MRPC}{multigap resistive plate chamber}
  \acro{EMC}{Electromagnetic Calorimeter}
  \acro{RPC}{Resistive Plate Chambers}
  \acro{FEE}{front-end electronics}

  \acro{HPTDC}{high performance time-to-digital converter}
  
 \end{acronym}

\end{document}